# On the Kramers-Kronig transform with logarithmic kernel for the reflection phase in the Drude model

*Jean-Michel André*[†,‡,*], *Karine Le Guen*[†,‡], *Philippe Jonnard*[†,‡],

*Nicola Mahne*[§], *Angelo Giglia*[§], *Stefano Nannarone*[§,¶]

(†) Laboratoire de Chimie Physique - Matière et Rayonnement, UPMC Univ Paris 06, 11 rue Pierre et Marie Curie, F-75231 Paris CEDEX 05, France

(‡) CNRS-UMR 7614, 11 rue Pierre et Marie Curie, F-75231 Paris CEDEX 05, France

(§) Laboratorio Nazionale TASC, INFM-CNR, s.s.14, km 163.5 in Area Science Park, I-34012 Trieste, Italy

(¶) Dipartimento di Ingegneria dei Materiali e dell Ambiente, Universita di Modena e Reggio Emilia, Via Vignolese 905, I-41100 Modena, Italy

**Short title**: Kramers-Kronig transform for the reflection phase in the Drude model

* corresponding author: jean-michel.andre1@upmc.fr ; phone: 33 1 44 27 66 14

We consider the Kramers-Kronig transform (KKT) with logarithmic kernel to obtain the reflection phase and subsequently the complex refractive index of a bulk mirror from reflectance. However, it remains some confusion on the formulation for this analysis. Assuming the damped Drude model for the dielectric constant and the oblique incidence case, we calculate the additional terms: phase at zero frequency and Blashke factor and we propose a reformulated KKT within this model. Absolute reflectance in the *s*-polarization case of a gold film is measured between 40 and 350 eV for various glancing angles using synchrotron radiation and its complex refractive index is deduced using the reformulated KKT that we propose. The results are discussed with respect to the data available in the literature.

## 1. Introduction

Kramers-Kronig transform (KKT) of the reflectance measurements remains a unique method to determine the dielectric constant of materials, mainly in the short wavelength domain (uv and soft x-ray ranges) [1] where other techniques such as ellipsometry are inadequate. In a large number of the early works [2], the KKT was implemented with a formulation in which the phase at an arbitrary frequency $\omega_0/2\pi$ is directly proportional to an integral with a logarithmic kernel over the entire spectrum :

$$\Psi(\omega_0) = -\frac{2\,\omega_0}{\pi} P\int_0^{+\infty} \frac{Ln[\rho(\omega)]}{\omega^2-\omega_0^2}\, d\omega \tag{1}$$

It is the Robinson-Price formula [3], where $\Psi$ and $\rho$ (assumed to be even) are the phase and the amplitude of the complex coefficient of reflection $r$, respectively and P stands for the principal-value.

Nevertheless it has been emphasized that it should be necessary to add to this integral a constant term, both in the context of the scattering theory [4,5] and in studies of the foundations of the dispersion relations, in particular by Toll [6]. Stern in a work based on Toll's analysis attempted to justify the KKT [7]. Young [8] discussed the KKT as given by Toll and Stern on the basis of physical considerations and concluded that the validity of the KKT remains in doubt and must be applied with caution. More recently Nash *et al*. [9] have derived what they called the « correct version » of the KKT for the phase spectrum with a logarithmic kernel and give :

$$\Psi(\omega_0) = \Psi(0) - \frac{2\,\omega_0}{\pi} P\int_0^{+\infty} \frac{Ln[\rho(\omega)]}{\omega^2-\omega_0^2}\, d\omega \tag{2}$$

Using integration by parts, Eq. (2) can be slightly rewritten as follows :

$$\Psi(\omega_0) = \Psi(0) - \frac{1}{\pi} P\int_0^{+\infty} \frac{dLn[\rho(\omega)]}{d\omega}\; Ln\left|\frac{\omega+\omega_0}{\omega-\omega_0}\right| d\omega \tag{3}$$

From the mathematical point of view, the term $\Psi(0)$ arises from the existence of a pole at zero frequency in the kernel that is overlooked in other approachs. Nash *et al*. [9] emphasized that the term $\Psi(0)$ is close to $-\pi$ for many materials, but not always, in particular for a material whose absorption coefficient does not vanish at zero frequency. Smith [10] derived a formula similar to Eq.(2) but proposed to eliminate the additional term $\Psi(0)$ for conductors and insulators. Lee *et al.* [11] have calculated the term $\Psi(0)$ assuming that the dielectric constant is given by the plasma model, *i.e.* the undamped Drude model, for the case of normal incidence using Eq. (3).

Otherwise it has been recognized that other terms, sometimes called Blashke factors, must be added; they correspond to the occurrence of singularities in the complex coefficient of reflection r along the imaginary frequency axis [12]. This complementary Blashke term has been formally expressed for any incidence as well as for the *s*- and *p*-polarization [13]. ~~For some cases,~~ Generally, the exact value of this term requires the knowledge of the behaviour of the dielectric constant along the imaginary frequency axis.

Thus, although KKT from reflectance measurements has been applied since many years to find the complex refractive index of materials, it seems that it remains some confusion on the formulation for this analysis. To the mere Robinson-Price formula usually implemented, it should be necessary to add some terms (zero-frequency phase term, Blashke factors) whose value depends on the behaviour of the dielectric constant and on the conditions of the measurements (oblique or normal incidence, polarization of the incident radiation, nature of the mirror: bulk, multilayer, ...). This work is a tentative to clarify this problem for any case of incidence in the framework of the Drude model; we only assume transverse electric polarization. We propose to treat the transverse magnetic in a forthcoming paper.

In the UV and soft x-ray domains, most of reflectance measurements are performed at oblique incidence for practical reasons [1] and consequently, an extension to the oblique incidence case of the works by Nash *et al.* [9] and Lee *et al.* [11] that were restricted to the normal incidence case, was needed. Thus we have been led to develop a reformulated KKT based on the assumption of the undamped Drude model for the dielectric constant.

To test the validity of our reformulation, we present a comparison of the values of the complex refractive index of a gold film deduced from our KKT analysis based on our own reflectance measurements (conditions of which are well mastered) with data from literature. The choice of the gold sample has been motivated by the fact that it is recognized that the tabulated values of the dielectric constant of this material obtained from methods different from KKT analysis by reflection are well established.

Initially this work has been motivated by the fact that the knowledge of accurate values of the dielectric constants in the short wavelength domain becomes very important for the development of new multilayer optics designed for this spectral domain [14] and for the characterization of new materials for microelectronics [15].

The paper is organized as follows. In section 2, we present a formulation of the KKT extended to the oblique case involving the phase at zero frequency and a Blashke factor. The phase at zero frequency is calculated within the Drude model in section 3, while the Blashke factor is determined in section 4. The final expression of the global phase in the Drude model is given in section 5. We explain in section 6 our method to deduce the complex refractive index. Section 7 is devoted to the presentation of our reflectance measurements of a gold film. In section 8 we present and discuss the results of the determination of the gold optical index from our measurement by means of our KKT analysis.

## 2. KKT formulation in normal and oblique case with the Drude model

It is well-known that from the mathematical point of view, the KKT can be regarded as a Hilbert transform that relates the real and imaginary part of a linear response function by virtue of the causality principle. The calculation of this Hilbert transform is performed by a contour integration on a judiciously chosen path C in the complex frequency $z$ plane with an appropriate kernel. The choice of the kernel and the determination of the contour C are strongly related. The strategy is to integrate a function that is analytic within the contour and on the contour C so that, by virtue of the Cauchy theorem, the contour integral vanishes. The path along the real axis must lead to a principal-value integral with a logarithmic kernel containing the modulus of the reflection coefficient and to the phase of reflection at the frequency $\omega_0$ of interest. These considerations lead to consider a kernel containing the term $\frac{Ln[r(z)]}{z-\omega_0}$. Consequently, in the usual KKT analysis the dispersion relationship is derived from the contour integral :

$$\oint \frac{Ln[r(z)]}{z-\omega_0} dz \qquad (4)$$

where the closed contour includes a complete semi-circle in the upper half-plane (UHP). To take into account the simple pole at $\omega = 0$, Nash *et al.* [9] used the kernel g($z$) Ln[r($z$)] with $g(z) = z^{-1} (z - \omega_0)^{-1}$ and an integration path $C_1$ (see Appendix) that consists of a complete semi-circle in the UHP and of small circles around the 0 and $\omega_0$ poles along the real axis; thus they derivate Eq. (2). Nevertheless Ln[r($z$)] may have singularities on the imaginary axis as discussed in [12] which compromise the analyticity of the kernel within the path $C_1$. These singularities appear along the imaginary frequency axis and have been given by Plaskett *et al.* [12] for the case where the reflection coefficient is given by the Fresnel formula. Taking into account all these considerations, it appears that the kernel f($z$) Ln[r($z$)] with $f(z) = (z^2 - \omega_0^2)^{-1}$ is analytic in the first quadrant of the UHP but has a pole at $z = \omega_0$ on the real frequency axis and possibly three poles on the imaginary frequency axis, commonly named $\tau$, $\eta$ and $\delta$ if the coefficient of reflection is assumed to be given by Fresnel formula.

Assuming that the radiation is specularly reflected in the *s*-polarization case (transverse electric) at a glancing angle $\theta$, the corresponding Fresnel coefficient of reflection is :

$$r_s = \frac{-\sin\theta + \sqrt{\varepsilon - \cos^2\theta}}{\sin\theta + \sqrt{\varepsilon - \cos^2\theta}} \tag{5}$$

The first singularity $\tau$ called pseudo total reflection frequency corresponds to the branch cut of the square root in the Fresnel formula Eq. (5) and satisfies :

$$\varepsilon(i\tau) = \cos^2\theta \tag{6}$$

The second one $\eta$ called pseudo zero total frequency corresponds to a divergence of the logarithm and fulfills :

$$\varepsilon(i\eta) = 1 \tag{7}$$

The third one $\delta$ called pseudo total polarization frequency corresponds to Brewster's condition for complete polarisation and satisfies :

$$\varepsilon(i\delta) = \cot\mathrm{an}^2\theta \tag{8}$$

To take into account these singularities one has to consider the closed contour $C_2$ (see Appendix). The calculation detailed in Appendix gives for the overall phase the following relationship:

$$\Psi(\omega_0,\theta) = \Psi(0,\theta) + K(\omega_0,\theta) + B(\omega_0,\theta) \tag{9}$$

The first term comes from the pole at $\omega = 0$ when one considers the kernel g($z$) Ln[r($z$)] and the path $C_1$ and does not occur with the kernel f($z$) Ln[r($z$)] and the path $C_2$; the second term $K(\omega_0,\theta)$ is a principal-value integral resulting from the integration along the real axis for both kernels and paths and is given by :

$$K(\omega_0,\theta) = -\frac{2\,\omega_0}{\pi} P\int_0^{+\infty} \frac{Ln[\rho(\omega)]}{\omega^2 - \omega_0^2} d\omega = -\frac{1}{\pi} P\int_0^{+\infty} \frac{dLn[\rho(\omega)]}{d\omega} Ln\left|\frac{\omega + \omega_0}{\omega - \omega_0}\right| d\omega \tag{10}$$

Eq. (10) is nothing else but the Robinson-Price formula. The third term is the the so-called Blashke factor arising from the integration along the imaginary axis for the kernel f($z$) Ln[r($z$)] and the path $C_2$ ; it does not exist for the kernel g($z$) Ln[r($z$)] and the path $C_1$. This term reads:

$$B(\omega_0,\theta) = -\frac{2\omega_0}{\pi} \int_0^{+\infty} \frac{\mathrm{Im}\left(Ln[\rho(i\omega)]\right)}{\omega^2 + \omega_0^2} d\omega \tag{11}$$

**3. Phase $\Psi(0,\theta)$ at origin**

In the Drude model, the dielectric constant is given by :

$$\varepsilon(\omega) = 1 - \chi + i\,\Gamma \tag{12a}$$

with
$$\chi = \frac{\omega_p^2}{\omega^2 + \gamma^2} \tag{12b}$$

and
$$\Gamma = \frac{\gamma\,\omega_p^2}{\omega^3 + \gamma^2 \omega} \tag{12c}$$

where $\omega_p$ is the plasma frequency and $\gamma$ the Drude relaxation rate. For gold $\omega_p$ is about 9 eV (1.37 $10^{16}$ rad/s) and $\gamma$ is around 35 meV (5 $10^{13}$ rad/s). The continuation of $\varepsilon(\omega)$ in the UHP ($\omega \to z = \omega' + i\,\omega''$) is analytic ; $\varepsilon$ takes real values only on the imaginary axis where it decreases from $\infty$ at $\omega''$ = i 0 to 1 at $\omega''$ = i $\infty$, as it can be seen from the following formula :

$$\varepsilon(i\omega'') = 1 + \frac{\omega_p^2}{\omega''^2 - \gamma^2} - \frac{\gamma\,\omega_p^2}{\omega''^3 - \gamma^2 \omega''} \tag{13}$$

By inserting Eq. (13) in the Fresnel formula Eq. (5), one gets for the amplitude of the coefficient of reflection in the *s*-polarization case:

$$\rho_{Drude}(\omega_0,\theta) = \frac{\sqrt{A(\omega_0,\theta) + C(\omega_0,\theta) - S^+(\omega_0,\theta,0) - S^-(\omega_0,\theta,0)}}{\sqrt{A(\omega_0,\theta) + C(\omega_0,\theta) + S^+(\omega_0,\theta,0) + S^-(\omega_0,\theta,0)}} \tag{14}$$

The phase in the same condition is given by :

$$\Psi_{Drude}(\omega_0,\theta) = \arctan\left[\frac{\sin\theta\;\; S^+(\omega_0,\theta,0)}{2\sqrt{\chi^2 + \left(\Gamma - \sin^2\theta\right)^2} - \frac{C(\theta)}{2}}\right] \tag{15}$$

where

$$A(\omega_0,\theta) = \sqrt{2}\,\sqrt{3 + 8\,(\Gamma^2 + \chi^2 - \Gamma) + 4\,(2\,\Gamma - 2)\cos(2\theta) + \cos(4\theta)} \tag{16a}$$

$$C(\theta) = 2\,(1 - \cos(2\theta)) \tag{16b}$$

$$S^{\pm}(\omega_0,\theta,\alpha) = 4\,\sqrt[4]{\chi^2 + \left(\Gamma - \sin^2\theta\right)^2}\,\sin\left[\alpha + \theta \pm \frac{1}{2}\arctan\left(\frac{\chi}{\left(\Gamma - \sin^2\theta\right)}\right)\right] \tag{16c}$$

The subscript "Drude" means that the corresponding quantities are calculated with the dielectric constant given by the Drude model. When the frequency tends towards zero, one finds that in the Drude model and for the *s*-polarisation case whatever the value of the glancing angle $\theta$ :

$$\Psi(0,\theta) = 0 \tag{17}$$

Let us mention that according to Lee *et al.* [11], in the normal incidence case :

$$\Psi(0,\frac{\pi}{2}) = -\pi \tag{18}$$

a result that does not agree with Eq. (17). In fact the value of $\Psi(0,\theta)$ as deduced by Lee *et al.* [11] is related to the value of $\Psi(\infty,\frac{\pi}{2})$ but the determination of the "exact" value of $\Psi(\infty,\theta)$ appears to be a delicate question discussed in particular by Young. In our Drude model we find that :

$$\Psi(\infty,\theta) = 0 \tag{19}$$

## 4. Blashke factor $B(\omega_0,\theta)$ in the Drude model

We suppose that the incident medium is vacuum and that the incident radiation is reflected in the *s*-polarization configuration; the coefficient of reflection is then given by the Fresnel formula, Eq. (5). To determine the Blashke factor, one has to calculate the integral given by Eq. (11). To do it this is valuable to note that in our case the singularity $\tau$ disappears at infinity since $\cos^2 \leq 1$ while the pole $\delta$ vanishes and $\eta$ tends towards infinity; indeed along the imaginary axis the dielectric constant tends to unity as the frequency goes to infinity as mentioned above which means that $\eta \to \infty$. It follows that the term Im(Log[r(iω'']) in the integrand is equal to $\pi$ on the imaginary positive axis and the integral becomes:

$$B(\omega_0,\theta) = -\frac{2\omega_0}{\pi}\int_0^{+\infty}\frac{\pi}{\omega^2+\omega_0^2}\,d\omega = -\frac{2\omega_0}{\pi}\frac{\pi^2}{2\omega_0} = -\pi \tag{20}$$

It appears that the Blashke factor contributes to the global phase by the term $-\pi$ in the Drude model, in agreement with the data given in [13], case 3.1 of the Table 1.

## 5. The global phase in the Drude model

By virtue of Eq. (9), the global phase in the Drude model reads:

$$\Psi_{Drude}(\omega_0,\theta) = \Psi_{Drude}(0,\theta) + K_{Drude}(\omega_0,\theta) + B_{Drude}(\omega_0,\theta) \tag{21}$$

Taking into account the preceding results, it follows that $\Psi_{Drude}(\omega_0,\theta)$ is given for the *s*-polarization reflection by :

$$\Psi_{Drude}(\omega_0,\theta) = -\frac{1}{\pi}\left(P\int_{\omega_{min}}^{\omega_{max}}\frac{dLn\left[\rho(\omega)\right]}{d\omega}\,Ln\left|\frac{\omega+\omega_0}{\omega-\omega_0}\right|d\omega\right) - \frac{1}{\pi}\left(P\int_{\omega_{max}}^{+\infty}\ldots + P\int_{0^+}^{\omega_{min}}\frac{dLn\left[\rho_{Drude}(\omega)\right]}{d\omega}\,Ln\left|\frac{\omega+\omega_0}{\omega-\omega_0}\right|d\omega\right) - \pi \tag{22}$$

The first term corresponds to the Robinson-Price term (R-P T) evaluated from the experimental amplitude of the coefficient of reflection measured in the interval ($\omega_{min}$, $\omega_{max}$); the second term corresponds to the R-P T evaluated from amplitude of the coefficient of reflection calculated outside the experimental domain: in high energy region according to the Drude model, one can show using Eq. (14) that $\rho(\omega)$ varies as $\omega^{-2}$ so that:

$$P\int_{\omega_{max}}^{+\infty} \frac{dLn[\rho_{Drude}(\omega)]}{d\omega} Ln\left|\frac{\omega+\omega_0}{\omega-\omega_0}\right| d\omega = -P\int_{\omega_{max}}^{+\infty} \frac{2}{\omega} Ln\left|\frac{\omega+\omega_0}{\omega-\omega_0}\right| d\omega \qquad (23)$$

In the low energy range the following extrapolation is adopted:

$$\rho(\omega) = \sqrt{\rho^2(\omega_{min}) + (1-\rho^2(\omega_{min}))\left(1-\frac{\omega}{\omega_{min}}\right)^{1/2}} \qquad (24)$$

from the point $\omega_{min}$ to $\rho = 1$ at $\omega = 0$.

The last term $-\pi$ of Eq. (23) is the Blashke factor. The phase at zero frequency does not contribute in our model. According to Eq. (5) the global phase could be obtained by the sum of only three terms; nevertheless other terms may be added. The term $m\,\pi\,\mathrm{sgn}(\omega)$ with $\mathrm{sgn}(\omega) = \frac{\omega}{|\omega|}$ for $\omega \neq 0$ must be added to take into account that the direction of beam propagation is arbitrary and the reflectivity tensor can be multiplied by -1 without affecting the physical situation. Of course, one may add an integer multiple of 2. Finally, let us emphasize that other terms should be introduced if the kernel presents at some frequencies a “pathological” behaviour compromising its analyticy in the UHP but this situation is beyond the scope of the paper since we restrict to the case where the coefficient of reflection is given by the Fresnel formula and the incident medium is vacuum.

**5. Method of determination of the dielectric constant by the KKT in the Drude model**

From the reflectance measurements, we deduce the experimental amplitude of the reflection coefficient that is then introduced in the integrand of the first integral of Eq. (22). One deduces an experimental coefficient of reflection:

$$r(\omega,\theta) = \rho_{exp}(\omega,\theta)\ \exp\left(i\,\Psi(\omega,\theta)\right) \qquad (25)$$

It is well-known that the surface roughness tends to decrease the specular reflectance of a mirror and this phenomenon must be taken into account. Although very sophisticated theories have been developed to model the light scattering by a surface roughness, we restrict ourselves to the so-called Debye-Waller (*DW*) model [16]. Let us recall that in this model the coefficient of reflection is decreased by the DW factor :

$$DW = \exp\left( -4\,\frac{\pi^2}{\lambda^2}\sin^2\theta\ \sigma^2 \right) \tag{26}$$

where λ is the radiation wavelength in the mirror medium and σ the *rms* height of the roughness.

To obtain the dielectric constant from the measurements we have implemented three methods :

-i/ the comparison of the theoretical coefficient $r_F(\omega,\theta,\varepsilon_r,\varepsilon_i)$ calculated by means of the Fresnel formula multiplied by the Debye-Waller term $DW(\theta,\omega)$ taking into account the surface roughness of the sample with the experimental coefficient by solving the following system of two equations :

$$\begin{aligned}\mathrm{Re}[r(\omega,\theta)] &= DW(\omega,\theta)\,\mathrm{Re}[r_F(\omega,\theta,\varepsilon_r,\varepsilon_i)]\\ \mathrm{Im}[r(\omega,\theta)] &= DW(\omega,\theta)\,\mathrm{Im}[r_F(\omega,\theta,\varepsilon_r,\varepsilon_i)]\end{aligned} \tag{27}$$

unknows of which are the real part $\varepsilon_r$ and the imaginary part $\varepsilon_i$ of the dielectric constant. The FindRoot function of Mathematica® was used to carry out this operation.

-ii/ the direct calculation of the real part $\varepsilon_r$ and of the imaginary part $\varepsilon_i$ of the dielectric constant from the experimental amplitude $\rho = \rho_{exp}(\omega,\theta)$ corrected by the DW term and the phase $\Psi_{Drude}(\omega,\theta)$ resulting from the KKT in the Drude model, by means of the relationships :

$$\varepsilon_r = 1 - \frac{4\,\rho\sin^2\theta\left(\cos\Psi\left(1+\rho^2\right)+2\,\rho\right)}{\left(1+\rho^2+2\,\rho\cos\Psi\right)^2} \tag{28}$$

$$\varepsilon_i = \frac{4\,\rho\sin^2\theta\ \sin\Psi\left(\rho^2-1\right)}{\left(1+\rho^2+2\,\rho\cos\Psi\right)^2} \tag{29}$$

-iii/ the direct calculation of the complex dielectric constant from the experimental amplitude $\rho = \rho_{\exp}(\omega,\theta)$ corrected by the DW term and from the phase $\Psi_{Drude}(\omega,\theta)$ by inverting the Fresnel formula :

$$\varepsilon = \cos^2\theta + \sin^2\theta\left(\frac{1-r(\omega,\theta)}{1+r(\omega,\theta)}\right)^2 \tag{30}$$

The real part of the optical index $n$ and the imaginary part $\beta$ are finally obtained from $\varepsilon_r$ and $\varepsilon_i$ using the well-known equations:

$$\varepsilon_r = n^2 - \beta^2 \tag{31}$$

$$\varepsilon_i = 2\, n\, \beta \tag{32}$$

## 6. Measurements of the reflectance of the gold mirror

The Au film was deposited using magnetron sputtering onto a $Si_3N_4$/Si substrate. The purity of the Au target is 99.95%. The thickness of the Au film is about 60 nm. Before the measurement the sample was cleaned by a soft Ar sputtering in the following conditions: 1 keV ion energy, sample orientation equal to 45°, base pressure around $2.10^{-7}$ mbar, duration 30 min. The contamination of the Au film was checked *in situ* by means of x-ray photoelectron spectroscopy and did not show C or O presence within the experimental error. The reflectance of the mirror was measured at the BEAR beam-line of ELETTRA synchrotron facility for *s*-polarization light. The surface of the sample was investigated with a ZYGO interferometer [17]. According to the location on the sample, the *rms* height of the surface $\sigma$ is between 1 and 5 nm. For the DW calculation we have retained the mean value of 1.5 nm.

Figure 1 shows the reflectance of the gold mirror for different glancing angles $\theta$: 5°, 10°, 30° 50° and 70°. Around 290eV and for large glancing angles, it is possible to note a signature of a carbon contamination of the sample; this contamination was not detected by XPS measurements and can be ascribed probably to buried carbon, maybe at the interface with the substrate: the carbon peaks are not visible at low glancing angles that is when the light doesn't reach the substrate. For $\theta = 50°$ and 70°, Kiessig fringes can be observed. It appears that the measurements at $\theta = 30°$ are well suited for the determination of the optical index because interferences with the substrate (Kiessig fringes) are not too strong and the penetration of the radiation in the sample is enough to probe the bulk of the Au film, which is not the case for the smallest glancing angles at energies smaller than 200 eV.

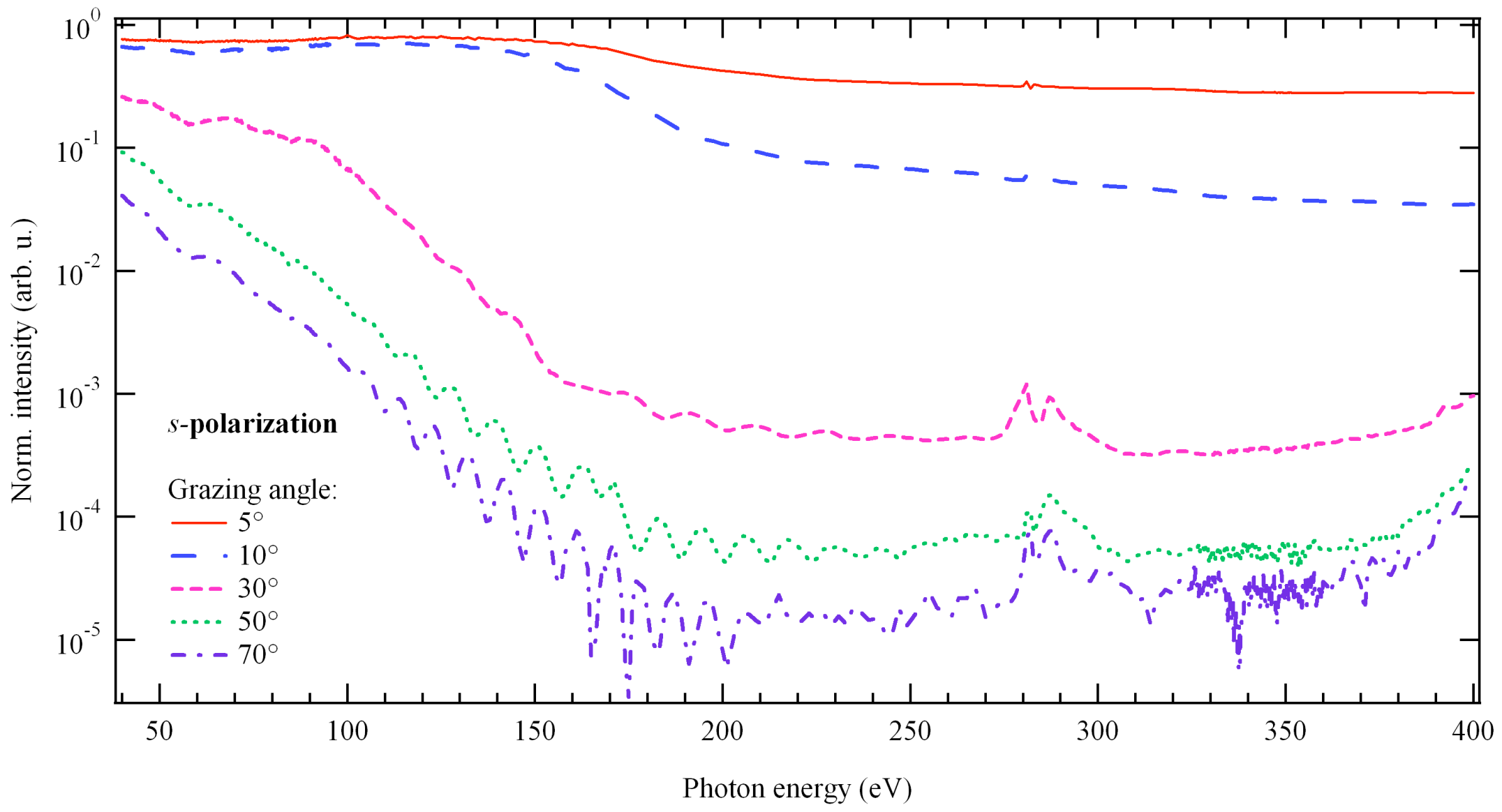

Figure 1 **:** Measured reflectance of the gold mirror for different glancing angles θ: 5°, 10°, 30° 50° and 70° in the *s*-polarization case.

## 7. Complex optical index of gold between 40 and 350 eV from KKT

The real part $n$ and imaginary part $\beta$ of the complex refractive index of our gold film have been determined as explained in section 5 from the measurements performed at θ = 30°. The three methods give the same results. Figure 2 presents the values of $n$ and $\beta$ between 40 and 350 eV. The effect of anomalous dispersion associated with the 5p and 4f electrons of gold can be observed around 60 and 85 eV respectively, as shown in Figure 2.

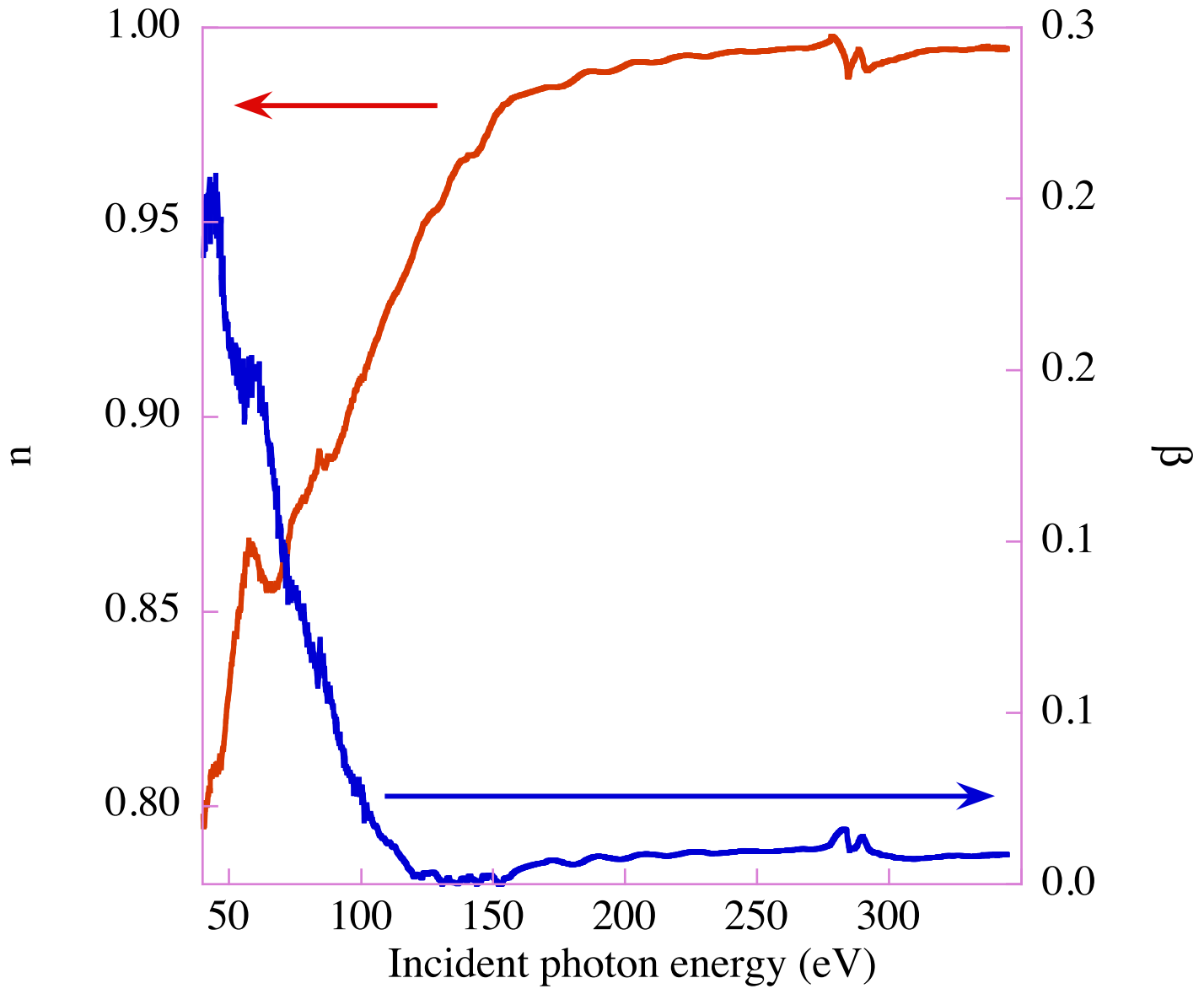

Figure 2: Real part $n$ (red line) and imaginary part $\beta$ (blue line) of the refractive index of our gold film from the measurements performed at θ = 30° calculated by our reformulated KKT.

The values obtained with our KKT can be compared to the tabulated values [18], see figure 3a for the real part $n$ and figure 3b for the imaginary part $\beta$. The agreement is rather satisfactory but it is difficult to draw conclusions from this comparison about the quality of the data because the methods of determination and the samples are rather different. Nevertheless one can say that our reformulated KKT with accurate measurements of reflectance and careful characterization of the surface in terms of roughness provides fairly correct values of the complex refractive index.

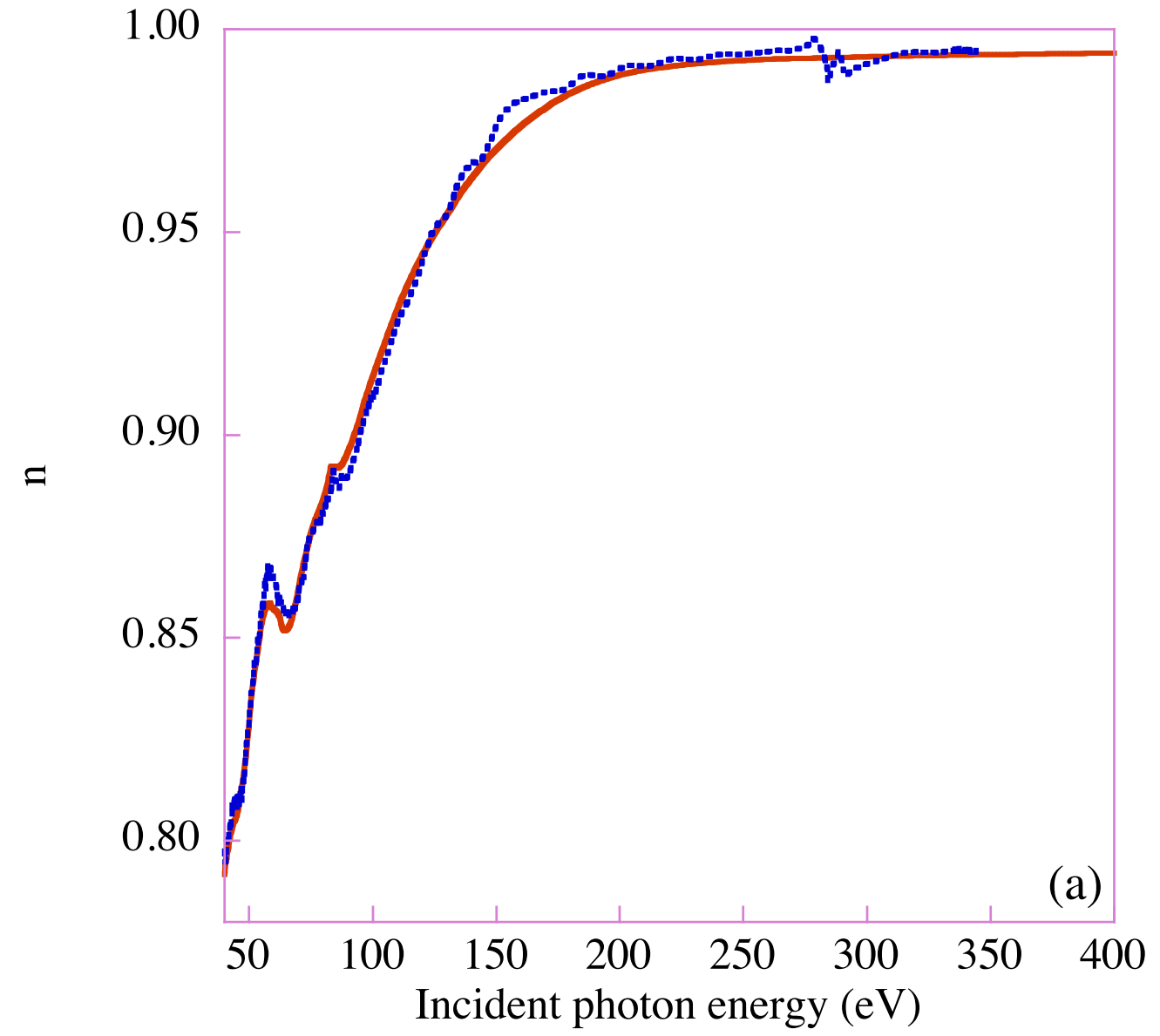

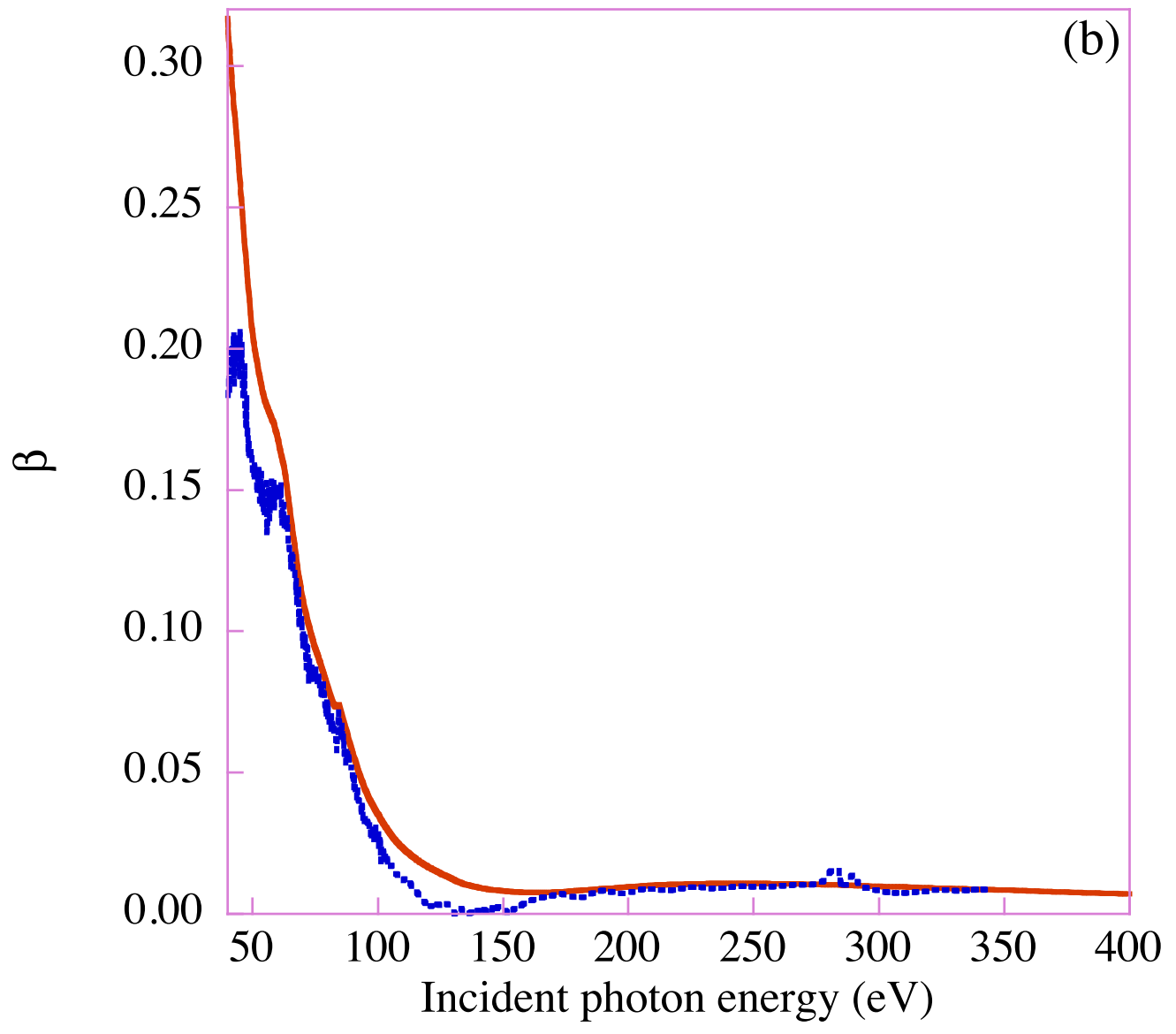

Figure 3: Real part $n$ (a) and imaginary part $\beta$ (b) of the refractive index of our gold film from the measurements performed at $\theta = 30°$, calculated by our reformulated KKT (blue line) compared to the tabulated values (red line).

## 8. Conclusion and perspectives

The approach proposed by Nash *et al*. [9] and Lee *et al*. [11] has been extended to the oblique case and to the Drude model -including the damping term- of the dielectric constant. In this framework we have calculated the additional term $\Psi(0,\theta)$ and the Blashke factor and finally deduced a reformulated KKT. To take into account the singularities on the imaginary frequency axis we have implemented a contour integration to perform the Hilbert transform different from the one chosen by Nash *et al*. [9] and we get a formulation of the KKT that is slightly different. It appears that in our approach the contribution to the global phase of the term $\Psi(0,\theta)$ cancels and that the Blashke term $B(\omega_0)$ contributes by the value $-\pi$.

Comparison of the results obtained for the complex refractive index of our gold sample by means our reformulated KKT with tabulated data leads to believe that our approach is a rather satisfactory method to determine the unknown refractive index values of newly elaborated substances provided that the reflecting surface is well characterized. It would be valuable to check this conclusion for other materials, especially semiconductors and insulators, the refractive indices of which are available in the literature. The case of *p*-polarization (transverse magnetic) especially concerning the influence of the Blashke factor in the Brewster condition will be considered in a forthcoming paper.

**Acknowledgments** : The authors wish to thank Prs. A. Maquet and E.O. Filatova for fruitful discussions and M. Hu for experimental help. G. Sostero of Sincrotrone Trieste is also thanked for the interferometry measurements.

**Appendix**

The Hilbert transformation is obtained from integration in the complex frequency $z$ = ω'+ iω" plane along the contour $C_1$ or $C_2$ displayed in Figure A1 and A2 respectively, according to the approach of interest.

In the first approach one has to calculate the integral **I** where a pole at $z = 0$ is considered:

$$\mathbf{I} = \oint_{C_1} \frac{S(z)}{z(z - \omega_0)} dz \qquad \text{(A1)}$$

where S(z) stands for $Ln[r(z)]$. Within the contour and on the contour $C_1$, the kernel is analytic and therefore **I** = 0. In the second approach one has to calculate the integral **J**:

$$\mathbf{J} = \oint_{C_2} \frac{S(z)}{(z^2 - \omega_0^2)} dz \qquad \text{(A2)}$$

Within the contour and on the contour $C_2$, the kernel is analytic and therefore **J** = 0.

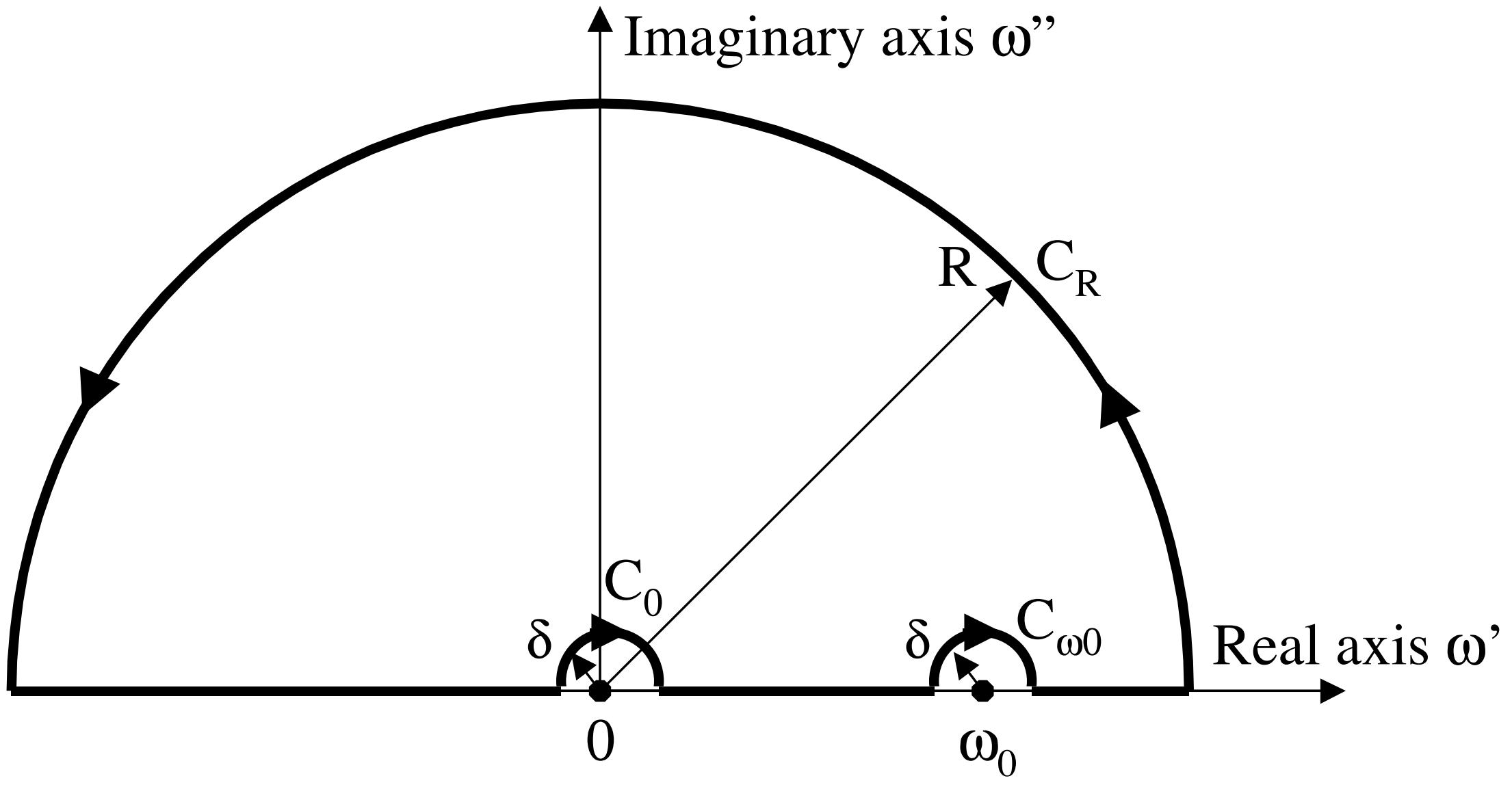

Figure A1: Integration contour $C_1$ (in bold line) in the complex frequency $z$ plane. The radius R of the part of circle $C_R$ may be made as large as necessary, and the radius δ of the semi-circles $C_0$ and $C_{\omega 0}$ centred at 0 and $\omega_0$ may be made as small as one pleases.

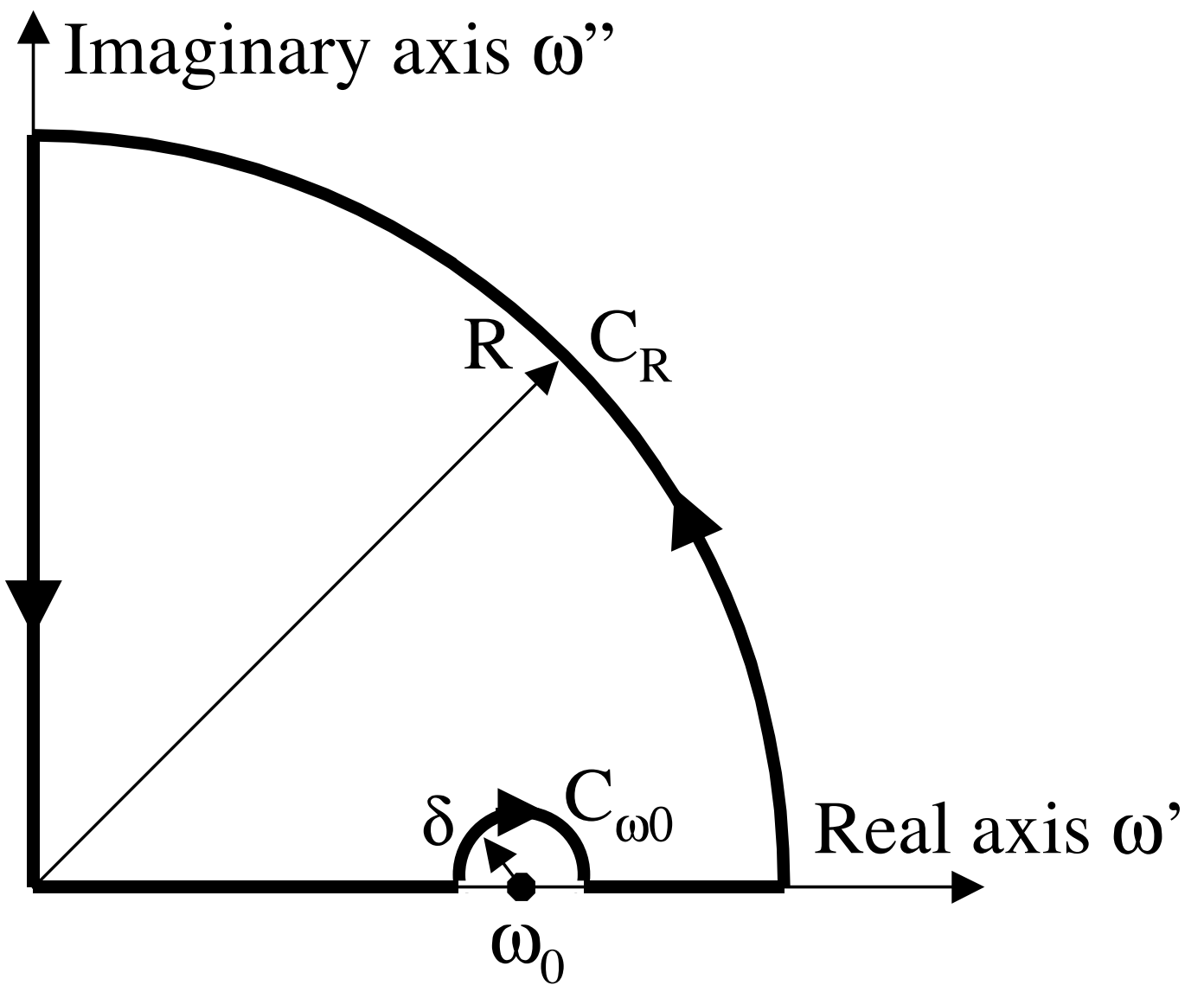

Figure A2: Integration contour $C_2$ (in bold line) in the complex frequency $z$ plane. The radius R of the part of circle $C_R$ may be made as large as necessary, and the radius δ of the semi-circle $C_{\omega 0}$ centred at $\omega_0$ may be made as small as one pleases.

The integrals **I** and **J** can be broken up as follows:

$$\mathbf{I} = \sum_i I_i \tag{A.2}$$

and

$$\mathbf{J} = \sum_i J_i \tag{A.3}$$

where

$$I_1 = \int_{C_R} \frac{S(z)}{z\left(z - \omega_0\right)} dz \,, I_2 = \int_{-R}^{-\delta} \frac{S(\omega')}{\omega' \left(\omega' - \omega_0\right)} d\omega' \,, I_3 = \int_{+\delta}^{-\delta+\omega_0} \frac{S(\omega')}{\omega' \left(\omega' - \omega_0\right)} d\omega \,,$$

$$I_4 = \int_{\delta+\omega_0}^{R} \frac{S(\omega')}{\omega' \left(\omega' - \omega_0\right)} d\omega' \,, I_5 = \int_{C_0} \frac{S(z)}{z\left(z - \omega_0\right)} dz \,,\ I_6 = \int_{C_{\omega 0}} \frac{S(z)}{z\left(z - \omega_0\right)} dz \tag{A.4}$$

and

$$J_1 = \int_{C_R} \frac{S(z)}{z^2 - \omega_0^2} dz \,,\ J_2 = -\int_R^0 \frac{S(\mathrm{i}\omega'')}{\omega''^2 + \omega_0^2} d\left(\mathrm{i}\omega''\right),\ \ J_3 = \int_0^{\omega_0 - \delta} \frac{S(\omega')}{\omega'^2 - \omega_0^2} d\omega' \,,$$

$$J_4 = \int_{C_{\omega 0}} \frac{S(z)}{z^2 - \omega_0^2} dz \,,\ J_5 = \int_{\omega_0 + \delta}^{R} \frac{S(\omega')}{\omega'^2 - \omega_0^2} d\omega' \tag{A.5}$$

Taking the limit of the integrals as $\mathrm{R} \to \infty$ and $\delta \to 0$, one gets:

$$I_1 \to 0 \,,\ J_1 \to 0 \tag{A. 6}$$

the integrals $I_1$ and $J_1$ along $C_R$ vanishes as the radius R goes to infinity, since the kernel remains finite in the UHP ; one has also :

$$I_2 + I_3 + I_4 \to P\int_{-\infty}^{+\infty} \frac{S(\omega')}{\omega'\left(\omega' - \omega_0\right)} d\omega' \qquad \text{(A.7)}$$

$$J_3 + J_5 \to P\int_{0}^{\infty} \frac{S(\omega')}{\omega'^2 - \omega_0^2} d\omega' \qquad \text{(A.8)}$$

The integral $\mathbf{I_5}$ along $C_0$ can be rewritten as:

$$I_5 = \varphi(\omega_0)\int_{C_0} \frac{dz}{z} + \int_{C_0} \frac{\varphi(z) - \varphi(0)}{z} dz \qquad \text{(A.9)}$$

where $\varphi(z) = \dfrac{S(z)}{z - \omega_0}$.

Setting $z = \delta \exp(i\alpha)$ in the first integral of Eq. (A.9) and integrating over $\alpha$ from $\pi$ to 0, leads to :

$$I_5 = -\, i\, \varphi(0)\, \pi + \int_{C_0} \frac{\varphi(z) - \varphi(0)}{z} dz \qquad \text{(A.10)}$$

Since $\varphi(z)$ is continuous at $z = 0$, which means that for all $\varepsilon > 0$ there exists a value $\xi$ such that if $|z| < \xi$, then $|\varphi(z) - \varphi(0)| < \varepsilon$, it follows that the absolute value of the integral in Eq. (A.10) satisfies :

$$\left|\int_{C_0} \frac{\varphi(z) - \varphi(0)}{z} dz\right| \le \frac{\varepsilon}{\xi}\left|\int_{C_0} dz\right| = \frac{\varepsilon}{\xi}\pi\sigma = \varepsilon \qquad \text{(A.11)}$$

which means that it can be made smaller than any pre-assigned number. Hence :

$$I_5 = i\,\frac{S(0)}{\omega_0}\pi \qquad \text{(A.12).}$$

A similar calculation gives for $I_6$ :

$$I_6 = -\, i\,\frac{S(\omega_0)}{\omega_0}\pi \qquad \text{(A.13)}$$

Collecting the preceding result gives :

$$\frac{\lim}{R \to \infty, \delta \to 0} I = P\int_{-\infty}^{+\infty} \frac{S(\omega')}{\omega'\left(\omega' - \omega_0\right)} d\omega' - \frac{i\pi}{\omega_0} S(\omega_0) + \frac{i\pi}{\omega} S(0) = 0 \qquad \text{(A.14)}$$

Taking the real part of Eq. (A.14) leads to:

$$\mathrm{Im}[S(\omega_0)] = \mathrm{Im}[S(0)] - \frac{2\omega_0}{\pi} P\int_{0}^{+\infty} \frac{\mathrm{Re}[S(\omega')]}{\omega'^2 - \omega_0^2} d\omega' \qquad \text{(A.15)}$$

Since:

$$S(\omega) = Ln[\mathrm{r}(\omega)] = Ln[\rho(\omega)\exp(i\,\Psi(\omega))] = Ln[\rho(\omega)] + i\Psi(\omega) \tag{A.16}$$

by inserting the real and imaginary parts, one gets:

$$\Psi(\omega_0) = \Psi(0) - \frac{2\omega_0}{\pi} P\int_0^{+\infty} \frac{Ln[\rho(\omega)]}{\omega^2 - \omega_0^2}\, d\omega \tag{A.17}$$

Similarly, collecting the $J_i$ integrals leads to:

$$\frac{\lim}{R\to\infty, \delta\to 0} J = P\int_0^{+\infty} \frac{S(\omega')}{\omega'^2 - \omega_0^2}\, d\omega' - \int_{+\infty}^{0} \frac{S(\mathrm{i}\omega'')}{\omega''^2 + \omega_0^2}\, d(\mathrm{i}\omega'') - \frac{i\pi}{2\omega_0} S(\omega_0) = 0 \tag{A.18}$$

Taking the real part of Eq. (A.18) leads to:

$$\mathrm{Im}[S(\omega_0)] = -\frac{2\omega_0}{\pi} P\int_0^{+\infty} \frac{Ln[\rho(\omega)]}{\omega^2 - \omega_0^2}\, d\omega - \frac{2\omega_0}{\pi}\int_0^{+\infty} \frac{\mathrm{Im}(S(\mathrm{i}\omega))}{\omega^2 + \omega_0^2} d\omega \tag{A.19}$$